\documentclass[12pt,thmsa]{article}
\usepackage{sw20lart}

\input{tcilatex}
\begin{document}

\author{Zhi Zhao$^{1}$, Jian-Wei Pan$^{2}$, and M.S. Zhan$^{1}$ \\
%EndAName
\textit{{\small $^{1}$ State Key Laboratory of Magnetic Resonance and
Atomic\ and Molecular Physics, }}\\
\textit{\small \ Wuhan Institute of Physics and Mathematics, the Chinese\
Academy of Sciences, } \\
\textit{\small Wuhan 430071, China }\\
\textit{{\small $^{2}$ Department of Mordern Physcis, }}\\
\textit{\small University of } \textit{\small Science and Technology of
China, }\\
\textit{\small Hefei 230027, China }}
\title{A Practical Scheme for Entanglement Concentration}
\date{}
\maketitle

\begin{abstract}
{\small We present a realistic purification scheme for pure non-maximally
entangled states. In the scheme, Alice and Bob at two distant parties first
start with two shared but less entangled photon pairs to produce a
conditional four-photon GHZ state, then perform a 45}$^{\circ }${\small \
polarization measurement onto one of the two photons at each party such that
the remaining two photons are projected onto a maximally entangled state. 
\newline
03.65.Bz, 03.67.-a, 42.50.Ar}
\end{abstract}

Quantum entanglement has become an important resource for quantum
computation \cite{dvi,ben1}, teleportation \cite{ben2}, dense coding \cite
{ben3}, and cryptography \cite{ek}. In the past few years, a large number of
experiments have shown that quantum computation and quantum communication
are more efficient in many aspects than their classical counterparts. In all
these experiments, maximally entangled states are usually required. However,
since there is decoherence during storage or transmission of particles over
noisy channels, the quality of entanglement is easily degraded. There are
two methods to overcome the effect of decoherence. One is the so-called
quantum error-correction scheme \cite{shor}, which makes quantum computation
possible despite the effect of decoherence and imperfect apparatus. The
alternative method is the entanglement purification. From the quantum
communication perspective, the entanglement purification is more powerful
than the quantum error correction. In order to achieve quantum communication
with high fidelity, the entanglement purification is necessary to obtain
maximally entangled states.

The basic idea of entanglement purification is to distill some pairs of
particles in highly entangled states from less entangled states using local
operations and classical communication. There have been several protocols 
\cite{ben4,ben5,deu,mu,br,cirac,bose} for purification of pure and mixed
non-maximally entangled states. In Schmidt-decomposition scheme \cite{ben4},
physical realization of local operations was achieved by collective
measurements. But, practically, it is very difficult to measure so many
photons simultaneously. Another similar scheme called Procrustean method 
\cite{ben4}, on the other hand, requires the states to be known in advance.
Entanglement purification schemes \cite{ben5,deu,mu,br,cirac} involving
quantum logic gates are even more difficult to implement for the mixed
states. The difficulties associated with different schemes block the way to
realize the purification experimentally. Recently, Bose et al. \cite{bose}
supposed that one could investigate the purification of entangled states via
entanglement swapping. By using Bell state measurements as local operations
and the measurement results as classical communication, such a purification
procedure could be easily realized by simple extension of an existing
entanglement swapping experiment \cite{pan1}. However, there one needs to
know the coefficients in advance in order to reconstruct the same entangled
states each time at Alice's or Bob's location.

In this paper, we present another protocol for entanglement concentration
based on the principle of quantum erasure \cite{scu1} and the Schmidt
projection method. In our scheme, one can concentrate entanglement from
arbitrary identical non-maximally entangled pairs at distant parties. For
non-maximally entangled states at distant parties we first erase the
``which-way'' information between the two non-maximally entangled states by
the process of quantum erasure such that we can produce a conditional
four-particle maximally entangled state. Then, after performing simple
Schmidt projection measurements \cite{peres} onto one of the two photons at
each parties, the remaining two photons are projected onto a maximally
entangled state. Hence, we provide a realistic scheme for the original
Schmidt-decomposition idea \cite{ben4}. On the other hand, we shall show
that our scheme can also be used to concentrate entanglement from
non-maximally entangled multi-photon states, for example, to distill a
Greenberger-Horne-Zeilinger (GHZ) state \cite{ac}.

Figure 1 is a schematic drawing of our purification scheme. Consider two
pairs of photons (1,2 ) and (3,4 ) in the following unknown polarization
entangled states: 
\begin{equation}
\ \left| \phi \right\rangle _{12}=\alpha \left| H_{1}\right\rangle \left|
H_{2}\right\rangle +\beta \left| V_{1}\right\rangle \left|
V_{2}\right\rangle ,
\end{equation}
\begin{equation}
\left| \phi \right\rangle _{34}=\alpha \left| H_{3}\right\rangle \left|
H_{4}\right\rangle +\beta \left| V_{3}\right\rangle \left|
V_{4}\right\rangle ,
\end{equation}
where $\left| \alpha \right| ^{2}+\left| \beta \right| ^{2}=1$, and Alice
holds photons 1 and 3, Bob holds photons 2 and 4. Our non-maximally
entangled state is the same as the one described in ref.[7].

Before proceeding to purify these states, the polarizations of photons 3 and
4 are rotated by 90$^{0}$ using two half-wave plates (HWP$_{90}$ in Fig. 1).
After passing through the two half-wave plates, the state of photons 3 and 4
becomes 
\begin{equation}
\left| \phi \right\rangle _{34}^{^{\prime }}=\alpha \left|
V_{3}\right\rangle \left| V_{4}\right\rangle +\beta \left|
H_{3}\right\rangle \left| H_{4}\right\rangle .\;
\end{equation}

Then we further forward photons 2 and 4 to a polarizing beam splitter PBS$%
_{B}$ (see Fig. 1). Suppose that photons 2 and 4 arrive at PBS$_{B}$
simultaneously such that the photons 2 and 4 interfere at the PBS$_{B}$.
Since the PBS transmits only the horizontal polarization component and
reflects the vertical component, after photons 2 and 4 passing through the
PBS$_{B}$ the total state of photons 1, 2, 3, and 4 evolves into 
\begin{eqnarray}
\left| \Psi \right\rangle &=&\alpha \beta \left| H_{1}\right\rangle \left|
H_{2^{^{\prime }}}\right\rangle \left| H_{3}\right\rangle \left|
H_{4^{^{\prime }}}\right\rangle +\alpha \beta \left| V_{1}\right\rangle
\left| V_{2^{^{\prime }}}\right\rangle \left| V_{3}\right\rangle \left|
V_{4^{^{\prime }}}\right\rangle  \nonumber \\
&&+\alpha ^{2}\left| H_{1}\right\rangle \left| V_{3}\right\rangle \left|
H_{4^{^{\prime }}}\right\rangle \left| V_{4^{^{\prime }}}\right\rangle
+\beta ^{2}\left| V_{1}\right\rangle \left| H_{2^{^{\prime }}}\right\rangle
\left| V_{2^{^{\prime }}}\right\rangle \left| H_{3}\right\rangle .
\end{eqnarray}

From the above equation, it is evident that Alice and Bob could observe a
four-fold coincidence among modes 1, 2$^{\prime }$ , 3, and 4$^{\prime }$
only for the terms $\left| H_{1}\right\rangle \left| H_{2^{^{\prime
}}}\right\rangle \left| H_{3}\right\rangle \left| H_{4^{^{\prime
}}}\right\rangle \;$ or $\left| V_{1}\right\rangle \left| V_{2^{^{\prime
}}}\right\rangle \left| V_{3}\right\rangle \left| V_{4^{^{\prime
}}}\right\rangle $. For the other two terms, there are always two particles
in one of the two output modes of the PBS$_{B}$ and no particle in the other
mode. Therefore, by only selecting those events there is exactly one photon
at the output mode $4^{^{\prime }}$ Alice and Bob can project the above
state into a maximally entangled four-particle state: 
\begin{equation}
\left| \Psi \right\rangle _{c}=\frac{1}{\sqrt{2}}\left[ \left|
H_{1}\right\rangle \left| H_{2^{^{\prime }}}\right\rangle \left|
H_{3}\right\rangle \left| H_{4^{^{\prime }}}\right\rangle +\left|
V_{1}\right\rangle \left| V_{2^{^{\prime }}}\right\rangle \left|
V_{3}\right\rangle \left| V_{4^{^{\prime }}}\right\rangle \right] ,
\end{equation}
with a probability of $2\left| \alpha \beta \right| ^{2}$.

Note that in the above description we have used the principle of quantum
erasure in a way that after PBS$_{B}$ some of the photons registered cannot
be identified anymore as to which source they came from. The PBS$_{B}$ plays
the double roles of both overlapping the two photons and erasing the
''which-way'' information. This principle, first proposed by Scully et al. 
\cite{scu1} and realized by many other authors \cite{scu2,kwiat,her,d01,kim1}%
, has been used in several important experiments such as quantum
teleportation \cite{b1}, entanglement swapping \cite{pan1}, three-particle
GHZ entanglement \cite{b2} and testing the non-locality of GHZ states \cite
{pan2}.

To generate maximally entangled two-photon state between Alice and Bob's
parties, they could further perform a 45$^{\circ}$ polarization measurement
onto the photons 3 and 4$^{\prime}$. As described in Fig. 1, Alice and Bob
first rotate the polarizations of the photon 3 and 4$^{\prime }$ by 45$^{0}$
with another two half-wave plates (refer to HWP$_{45}$ in Fig. 1). The
unitary transformation of the photons 3 and 4$^{\prime }$ through the
half-wave plates is given by

\begin{equation}
\left| H_{3}\right\rangle \rightarrow \frac{1}{\sqrt{2}}\left( \left|
H_{3}\right\rangle +\left| V_{3}\right\rangle \right)
\end{equation}

\begin{equation}
\left| V_{3}\right\rangle \rightarrow \frac{1}{\sqrt{2}}\left( \left|
H_{3}\right\rangle -\left| V_{3}\right\rangle \right)
\end{equation}

\begin{equation}
\left| H_{4^{\prime }}\right\rangle \rightarrow \frac{1}{\sqrt{2}}\left(
\left| H_{4^{\prime }}\right\rangle +\left| V_{4^{\prime }}\right\rangle
\right)
\end{equation}

\begin{equation}
\left| V_{4^{\prime }}\right\rangle \rightarrow \frac{1}{\sqrt{2}}\left(
\left| H_{4^{\prime }}\right\rangle -\left| V_{4^{\prime }}\right\rangle
\right)
\end{equation}

After this operation, the state (5) will evolve into a coherent
superposition of the following four combinations

\begin{eqnarray}
\frac{1}{2\sqrt{2}}\left| H_{3}\right\rangle \left| H_{4^{\prime
}}\right\rangle \left( \left| H_{1}\right\rangle \left| H_{2^{\prime
}}\right\rangle +\left| V_{1}\right\rangle \left| V_{2^{\prime
}}\right\rangle \right) + \\
\frac{1}{2\sqrt{2}}\left| V_{3}\right\rangle \left| V_{4^{\prime
}}\right\rangle \left( \left| H_{1}\right\rangle \left| H_{2^{\prime
}}\right\rangle +\left| V_{1}\right\rangle \left| V_{2^{\prime
}}\right\rangle \right) + \\
\frac{1}{2\sqrt{2}}\left| H_{3}\right\rangle \left| V_{4^{\prime
}}\right\rangle \left( \left| H_{1}\right\rangle \left| H_{2^{\prime
}}\right\rangle -\left| V_{1}\right\rangle \left| V_{2^{\prime
}}\right\rangle \right)+ \\
\frac{1}{2\sqrt{2}}\left| V_{3}\right\rangle \left| H_{4^{\prime
}}\right\rangle \left( \left| H_{1}\right\rangle \left| H_{2^{\prime
}}\right\rangle -\left| V_{1}\right\rangle \left| V_{2^{\prime
}}\right\rangle \right) .
\end{eqnarray}

Now, Alice and Bob let the photons 3 and 4$^{\prime }$ pass through the
polarization beam splitters PBS$_{3}$ and PBS$_{4^{\prime }}$ respectively,
and observe the coincidence between either detectors D$_{H3}$ and D$%
_{H4^{\prime }}$, or D$_{V3}$ and D$_{V4^{\prime }}$, or D$_{H3}$ and D$%
_{V4^{\prime }}$, or D$_{V3}$ and D$_{H4^{\prime }}$. Apparently, Alice and
Bob will observe four possible coincidences, i.e.$\left| H_{3}\right\rangle
\left| H_{4^{\prime }}\right\rangle$, $\left| V_{3}\right\rangle \left|
V_{4^{\prime }}\right\rangle $, $\left| H_{3}\right\rangle \left|
V_{4^{\prime }}\right\rangle$ and $\left| V_{3}\right\rangle \left|
H_{4^{\prime }}\right\rangle$. Following eq. (10), if both photons 3 and 4$%
^{\prime }$ are observed to be in the same polarization state (either $%
\left| H_{3}\right\rangle \left| H_{4^{\prime }}\right\rangle$ or $\left|
V_{3}\right\rangle \left| V_{4^{\prime }}\right\rangle $), then the
remaining two photons 1 and 2$^{\prime }$ are left in the state

\begin{equation}
\left|\phi^{+}\right\rangle_{12^{\prime}}=\frac{1}{\sqrt{2}}\left( \left|
H_{1}\right\rangle \left| H_{2^{\prime }}\right\rangle +\left|
V_{1}\right\rangle \left| V_{2^{\prime }}\right\rangle \right),
\end{equation}

Similarly, if photons 3 and 4$^{\prime }$ are observed to be in different
polarization state (either $\left| H_{3}\right\rangle \left| V_{4^{\prime
}}\right\rangle$ or $\left| V_{3}\right\rangle \left| H_{4^{\prime
}}\right\rangle $), then the remaining two photons 1 and 2$^{\prime }$ are
left in the state

\begin{equation}
\left|\phi^{-}\right\rangle_{12^{\prime}}=\frac{1}{\sqrt{2}}\left( \left|
H_{1}\right\rangle \left| H_{2^{\prime }}\right\rangle -\left|
V_{1}\right\rangle \left| V_{2^{\prime }}\right\rangle \right)
\end{equation}

In order to generate the same state $\left|\phi^{+}\right\rangle_{12^{%
\prime}}$ at each successful run, either Alice or Bob could perform an
additional local operation, i.e. a 180$^{0}$ phase shift (not shown in Fig.
1), to transform the state $\left|\phi^{-}\right\rangle_{12^{\prime}}$ into $%
\left|\phi^{+}\right\rangle_{12^{\prime}}$, conditioned upon that the
photons 3 and 4$^{\prime }$ are observed to be in different polarization
state. After performing the polarization measurements and the conditional
local operation, Alice and Bob can thus generate the maximally entangled
state $\left|\phi^{+}\right\rangle_{12^{\prime}}$ with a probability of 2$%
\left| \alpha \beta \right| ^{2}$, which is equal to the probability to
obtain the state $\left| \Psi \right\rangle _{c}$.

Here it is worthwhile to note that a detection of four-fold coincidence is
not necessary. In practice, with the help of single-photon detector \cite
{kim2} it is sufficient to measure the photon number and their polarizations
in 45$^{\circ }$ basis at the output modes 3 and 4$^{\prime }$. Conditioned
upon detecting exactly one photon at each of the two output modes 3 and 4$%
^{\prime }$, the remaining two photons 1 and $2^{\prime }$ can be prepared
in the state $\left| \phi ^{+}\right\rangle _{12^{\prime }}$ for further
application.

Furthermore, one can easily verify that our scheme can also be used to
concentrate entanglement from nonmaximally entangled multi-particle states.
Let us, for example, consider Alice, Bob and Cliff share two pairs of
nonmaximally entangled three-photon states $\alpha \left| H\right\rangle
\left| H\right\rangle \left| H\right\rangle +\beta \left| V\right\rangle
\left| V\right\rangle \left| V\right\rangle $ at three distant parties.
Through the similar process of quantum erasure and a 45$^{\circ}$
polarization measurement one of the two photons at each party, Alice, Bob
and Cliff would first share a conditional six-particle maximally entangled
state, and then obtain a maximally entangled state, i.e., GHZ entanglement
among three distant parties. The probability to obtain the GHZ state is
again 2$\left| \alpha \beta \right| ^{2}$.

Just as in many other schemes \cite{ben4,bose}, while entanglement of some
particles is concentrated by sacrificing the entanglement of other
particles, our procedure is only restricted to purification of two identical
non-maximally entangled states. Also, it should be noted that our scheme is
not optimal since the whole amount of entanglement is decreased by a factor
of 2 after finishing our purification procedure. However, our scheme is not
involved in collective measurement like the Schmidt-decomposition scheme 
\cite{ben4} and do not require the states to be known in advance like the
Procrustean method \cite{ben4}. Since the technique developed in the
experiments on quantum teleportation \cite{pan1,b1} and multi-photon
entanglement \cite{b2,pan2}, our scheme is within the reach of current
technology and thus is a feasible one of the original Schmidt-decomposition
scheme\cite{ben4}.

In summary, we present a practical scheme for purification of pure
non-maximally entangled states based on the principle of quantum erasure and
the Schmidt projection measurement. Using this scheme we can obtain
maximally entangled pairs, i.e., Bell states by a simple 45$^{\circ }$
polarization measurement. Our scheme might be useful in future long-distance
quantum communication.

\textsl{Note added}

During preparation of this manuscript, the authors became aware of related
work by Yamamoto et al., who arrive at the same proposal \cite{Imoto1}.
Also, a more powerful purification scheme working for general mixed
entangled states has been proposed recently by Pan et al. \cite{pan3}.

This work was supported by the National Natural Science Foundation of China.

* Present address: Institute for Experimental Physics, Boltzmanngasse 5,
University of Vienna, 1090 Austria.

\newpage
figure caption \newline
A schematic drawing of our scheme for entanglement concentration.
PBS$_{B}$, PBS$_{3}$, and PBS$_{4^{\prime }}$ are three polarization beam
splitters, which transmit the horizontal polarization component and reflect
the vertical component. The half-wave plates HWP$_{45}$ and HWP$_{90}$
rotate the horizontal and vertical polarization by $45^{0}$ and $90^{0}$
respectively; $D_{H3}$, $D_{V3}$, $D_{H4^{\prime }}$, and $D_{V4^{\prime }}$
are four single-photon detectors.

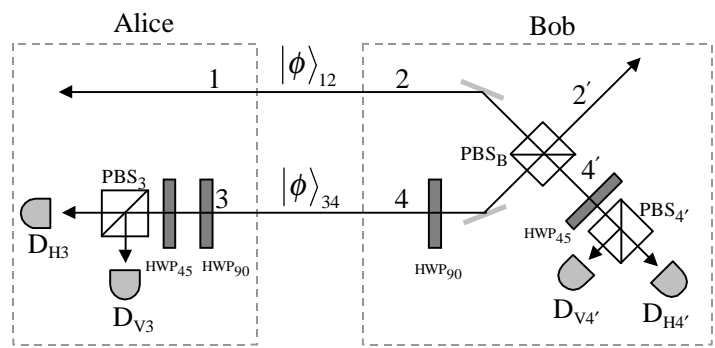
\end{document}